\documentclass[sigconf]{acmart}

\usepackage{textcomp}
\usepackage{stfloats}
\usepackage{url}
\usepackage{verbatim}
\usepackage{amsthm}
\usepackage{subfigure}
\usepackage{makecell}
\usepackage{multirow}
\usepackage{enumitem}
\usepackage{subcaption}
\usepackage{algorithm}
\usepackage{algorithmic}

\usepackage{color, colortbl}

\definecolor{Gray}{rgb}{0.9, 0.9, 0.9}

\def\model{HPMRec}
\usepackage{float}

\AtBeginDocument{%
  }

\copyrightyear{2025}
\acmYear{2025}
\setcopyright{cc}
\setcctype{by}
\acmConference[CIKM '25]{Proceedings of the 34th ACM International
Conference on Information and Knowledge Management}{November 10--14,
2025}{Seoul, Republic of Korea}
\acmBooktitle{Proceedings of the 34th ACM International Conference on
Information and Knowledge Management (CIKM '25), November 10--14, 2025,
Seoul, Republic of Korea}\acmDOI{10.1145/3746252.3761174}
\acmISBN{979-8-4007-2040-6/2025/11}

\begin{document}

\title[HPMRec]{Hypercomplex Prompt-aware Multimodal Recommendation}



\author{Zheyu Chen}
\authornote{Equal contribution}
\affiliation{%
  \institution{{The Hong Kong Polytechnic University}}
  \city{Hong Kong SAR}
  \country{China}}
\email{zheyu.chen@connect.polyu.hk}

\author{Jinfeng Xu}
\authornotemark[1]
\affiliation{%
  \institution{{The University of Hong Kong}}
  \city{Hong Kong SAR}
  \country{China}}
\email{jinfeng@connect.hku.hk}

\author{Hewei Wang}
\affiliation{%
  \institution{{Carnegie Mellon University}}
  \city{Pittsburgh}
  \state{PA}
  \country{USA}}
\email{heweiw@andrew.cmu.edu}

\author{Shuo Yang}
\affiliation{%
  \institution{{The University of Hong Kong}}
  \city{Hong Kong SAR}
  \country{China}}
\email{shuoyang.ee@gmail.com}

\author{Zitong Wan}
\affiliation{%
  \institution{{University College Dublin}}
  \city{Dublin}
  \country{Ireland}}
\email{zitong.wan@ucdconnect.ie}

\author{Haibo Hu}
\authornote{Corresponding author}
\affiliation{%
  \institution{{The Hong Kong Polytechnic University}}
  \city{Hong Kong SAR}
  \country{China}}
\email{haibo.hu@polyu.edu.hk}

\renewcommand{\shortauthors}{Zheyu Chen et al.}


\begin{abstract}
Modern recommender systems face critical challenges in handling information overload while addressing the inherent limitations of multimodal representation learning. Existing methods suffer from three fundamental limitations: (1) restricted ability to represent rich multimodal features through a single representation, (2) existing linear modality fusion strategies ignore the deep nonlinear correlations between modalities, and (3) static optimization methods failing to dynamically mitigate the over-smoothing problem in graph convolutional network (GCN). To overcome these limitations, we propose \textbf{HPMRec}, a novel \textbf{H}ypercomplex \textbf{P}rompt-aware \textbf{M}ultimodal \textbf{Rec}ommendation framework, which utilizes hypercomplex embeddings in the form of multi-components to enhance the representation diversity of multimodal features. HPMRec adopts the hypercomplex multiplication to naturally establish nonlinear cross-modality interactions to bridge semantic gaps, which is beneficial to explore the cross-modality features. HPMRec also introduces the prompt-aware compensation mechanism to aid the misalignment between components and modality-specific features loss, and this mechanism fundamentally alleviates the over-smoothing problem. It further designs self-supervised learning tasks that enhance representation diversity and align different modalities. Extensive experiments on four public datasets show that HPMRec achieves state-of-the-art recommendation performance.
\end{abstract}



\maketitle

\section{Introduction}

In the context of the exponential expansion of data volume, users encounter a significant challenge of information overload, thereby rendering recommender systems an effective method to mitigate this problem \cite{he2020lightgcn, xu2024aligngroup, xu2024fourierkan}. In multimodal recommendation scenarios, the synergy effect across modalities effectively mitigates the inherent data sparsity problem in traditional recommender systems\cite{xu2025mdvt, xu2025nlgcl}. Since the data structure of the recommender system has a natural bipartite graph structure, graph convolutional network (GCN) technology is also widely used in the recommender system \cite{chen2025don, xu2025cohesion, chen2025squeeze, xu2025best}. Recently, the design of representation learning using hypercomplex algebra has garnered interest, and several works \cite{chen2021quaternion, li2020quaternion, tran2020quaternion, zhang2019quaternion} have started to investigate hypercomplex-based recommender systems in the conventional recommendation field. This design shares a similar motivation with multi-head mechanisms, as it enables parallel learning of diverse representations. Therefore, hypercomplex algebras offer a more expressive mathematical framework and enhance the capacity to encode multimodal information.

Despite these works exploring the multimodal recommendation and hypercomplex embedding ability, they still face three fundamental limitations:
\noindent \textbf{Limitation 1:} Due to the richness of multimodal information, it is not sufficient for a single embedding to fully describe a user/item for each modality, and the traditional embedding structure restricts the representation diversity of multimodal features. 
\noindent \textbf{Limitation 2:} Existing linear modality fusion strategies (weighted sums or concatenations) ignore deep nonlinear correlations between modalities, which makes it difficult to fully explore the latent relationship between modalities, and leads to suboptimal exploration of cross-modality features.
\noindent \textbf{Limitation 3:} For the over-smoothing problem, which indicates the representation tends to be indistinguishable from those of their neighbors during the message passing in GCN, existing methods \cite{zhou2023layer,mao2021ultragcn,liu2021interest} manually design static optimization strategies to mitigate the over-smoothing problem, without considering the dynamic mechanism.

To overcome these limitations, we propose the \textbf{H}ypercomplex \textbf{P}rompt-aware \textbf{M}ultimodal \textbf{Rec}ommendation \textbf{(HPMRec)} framework with the following tailored designs. 
Inspired by \cite{li2022hypercomplex}, we introduce the Cayley-Dickson algebra, an elegant structure of hypercomplex embedding that contains multiple components, as the structure of each modality's user/item representation. Based on this structure, we propose a hypercomplex graph convolution operator that learns these representations, enabling each component to capture diverse modality-specific features.
Secondly, instead of regular dot products, the hypercomplex multiplication captures latent relations between two embeddings' components. We adopt this multiplication between different modalities' representations to capture nonlinear relations, which is beneficial to mine cross-modality features.
In addition, we introduce a learnable prompt to dynamically compensate for the misalignment of components and the core modality-specific features loss. Moreover, the diversity of representations mitigates the over-smoothing problem by ensuring that the representations of users and items remain distinguishable from those of their neighbors.
Furthermore, we design two self-supervised learning tasks. Specifically, we enhance user/item representation diversity by expanding the discrepancy between different components in hypercomplex embeddings, and we adopt cross-modality alignment, which also benefits modality fusion.

To summarize, our contributions are highlighted as follows:
\begin{itemize}[leftmargin=*]
    \item We propose the \textbf{HPMRec}, which utilizes hypercomplex embedding in the form of multi-components to enhance the representation diversity of modality-specific features. HPMRec leverages a novel nonlinear fusion strategy based on hypercomplex multiplication to bridge the semantic gap between modalities. 

    \item We design the prompt-aware compensation mechanism to dynamically compensate for component misalignment and core modality-specific feature loss. This module also alleviates the over-smoothing problem.
    
    \item Our HPMRec integrates self-supervised learning tasks to enhance modal representation diversity by expanding the discrepancy between components to enhance the diversity of representation, and we also implement cross-modality alignment, which is beneficial for modality fusion.

    \item We conduct comprehensive experiments to show the effectiveness and robustness of HPMRec. These results show that our HPMRec outperforms state-of-the-art methods.
\end{itemize}
\section{Related Work}
In this section, we will introduce the latest works in multimodal recommendation methods, the application of hypercomplex algebra in recommendation systems, and the development of prompts in recommendation systems.

\subsection{Multimodal Recommendation}
To mitigate the data sparsity problem, recent multimodal recommendation models leverage visual and textual features through matrix factorization \cite{he2016vbpr,chen2019personalized} and graph-based architectures \cite{wei2019mmgcn,wang2021dualgnn,zhou2023tale}. However, despite performance gains, three critical limitations remain. First, traditional embedding structures often force rich multimodal semantics into fixed-dimensional representations, hindering expressiveness. Second, modality fusion is typically handled via early \cite{he2016vbpr,zhang2021mining} or late \cite{wei2019mmgcn,wang2021dualgnn} strategies, both relying on linear operations that fail to capture latent cross-modality relations. Third, GCN-based methods such as NGCF \cite{wang2019neural} and LightGCN \cite{he2020lightgcn} suffer from over-smoothing, which recent multimodal extensions \cite{mao2021ultragcn, zhou2023layer} only mitigate through a static optimization strategy, lacking the ability of dynamic compensation. To this end, we propose the HPMRec, a novel framework that can overcome these limitations.

\subsection{Hypercomplex-based Recommendation}
Hypercomplex-based representation learning has proven effective in domains like computer vision \cite{zhu2018quaternion} and natural language processing \cite{tay2019lightweight}. More recently, researchers have begun applying these techniques to recommender systems: previous works \cite{li2022hypercomplex, chen2022breaking} focused on pure collaborative filtering using interaction data, while subsequent studies \cite{chen2021quaternion, li2020quaternion, tran2020quaternion} augmented that foundation by integrating auxiliary side information. Nevertheless, the inherent multi-component structure of the hypercomplex embedding makes it particularly suitable for encoding complex information such as multimodal features. To the best of my knowledge, no previous work has leveraged hypercomplex embeddings for multimodal recommendation. Our proposed HPMRec framework fills this gap and explores how hypercomplex embedding can benefit multimodal features through the representation capacity and structure.

\subsection{Prompt-based Recommendation}
Prompt learning has become an emerging research direction in the context of large pretrained models \cite{brown2020language, liu2023pre}, and some works explore the ability of prompt learning in the recommendation field. GraphPrompt \cite{liu2023graphprompt} defines the paradigm of prompts on graphs. To transfer knowledge graph semantics into task data, KGTransformer \cite{zhang2023structure} regards task data as a triple prompt for tuning. Additionally, prompt-based learning has also been introduced to enhance model fairness \cite{wu2022selective}, sequence learning \cite{xin2022rethinking}. Recently, PromptMM \cite{wei2024promptmm} proposed a novel multimodal prompt learning method that can adaptively guide knowledge distillation. In our HPMRec, we consider using the prompt's capabilities to implement a dynamic compensation mechanism of the hypercomplex embedding, so that it can achieve diverse representations while retaining core modality-specific features. It also alleviates the inherent over-smoothing problem of graph convolutional networks through the diversity of representations. This design fully and reasonably utilizes the prompt's capabilities to improve recommendation performance.

\section{Preliminary}
\subsection{Hypercomplex Algebra}
A hypercomplex number $h_x \in \mathbb{H}_n$ in $n$-dimensional real vector space can be expressed as a representation in the form as follows:
\begin{align}
    h_x=x_1\mathbf{i}_1 + x_2\mathbf{i}_2 + \cdots + x_n\mathbf{i}_n = \sum_{k=1}^{n} x_k\mathbf{i}_k,
    \label{eq:1}
\end{align}
where $x_1, x_2, \cdots, x_n$ denote distinct components of the hypercomplex number. The elements $\mathbf{i}_1, \mathbf{i}_2, \cdots, \mathbf{i}_n$ are called hyperimaginary units, where $\mathbf{i}_1=1$ represents the vector identity element \cite{alfsmann2006families}.

\subsection{Cayley–Dickson Construction}
The Cayley–Dickson algebra $\mathcal{A}$ is a sequence of hypercomplex algebras constructed from the real numbers using the Cayley–Dickson construction \cite{baez2002octonions, dickson1919quaternions}. Higher-dimensional Cayley–Dickson algebras can be obtained by doubling smaller algebras within the Cayley–Dickson construction \cite{lazendic2018hypercomplex}. Thus, the dimension of these algebras is a power of two. Specifically, such a construction procedure utilizes the $n$-th algebra $\mathcal{A}_n \in \mathbb{H}_{2^n}$ in the sequence to define the (n+1)-th algebra $\mathcal{A}_{n+1} \in \mathbb{H}_{2^{n+1}}$ as follows:
\begin{align}
    \mathcal{A}_{n+1} = \{h_a + h_b\mathbf{i}_{2^n+1}\},\quad \text{and}\ \ h_a, h_b \in \mathcal{A}_n,
    \label{eq:2}
\end{align}
where $n \in \mathbb{N}$ and $\mathcal{A}_0 = \mathbb{R}$. Here $\mathbf{i}_{2^n+1} \notin \mathcal{A}_n$ is the additional hyperimaginary unit for doubling the dimension of $\mathcal{A}_n$, satisfying the following rules: $(\mathbf{i}_{2^n+1})^2=-1$, $\mathbf{i}_{1}\mathbf{i}_{2^n+1}=\mathbf{i}_{2^n+1}\mathbf{i}_{1}$ and $\mathbf{i}_{o}\mathbf{i}_{2^n+1}=-\mathbf{i}_{2^n+1}\mathbf{i}_{o}=\mathbf{i}_{2^n}\mathbf{i}_{o}$ for all $o=2, 3, \cdots, 2^{n}$.

The mathematical operations for Cayley-Dickson algebras are defined recursively \cite{baez2002octonions, culbert2007cayley, dickson1919quaternions}. For $h_x=h_a+h_b\mathbf{i}_{2^n+1} \in \mathcal{A}_{n+1}$, $h_y=h_c+h_d\mathbf{i}_{2^n+1} \in \mathcal{A}_{n+1};$ $h_a \in \mathcal{A}_{n},\ h_b \in \mathcal{A}_{n},\ h_c \in \mathcal{A}_{n},\ h_d \in \mathcal{A}_{n}$; and $\gamma \in \mathbb{R}$, several widely utilized operations for Cayley-Dickson algebras are introduced as follows:
\begin{itemize}[leftmargin=*]
    \item The \textbf{addition} of $h_x$ and $h_y$ is defined as: $h_{x} \oplus_{\mathrm{n}+1} h_{y} = \left(h_{a} \oplus_{\mathrm{n}} h_{c}\right)+\left(h_{b} \oplus_{\mathrm{n}} h_{d}\right) \mathbf{i}_{2^{n}+1}.$ The \textit{subtraction} follows the same principle analogously, but flipping $\oplus$ with $\ominus$.
    
    \item The \textbf{conjugate} of $h_x$ is defined as: $\bar{h}_x = \bar{h}_a-h_b\mathbf{i}_{2^n+1}.$
    The conjugation for every $a \in \mathbb{R}$ is defined as: $\bar{a}=a$. $h_{x} \oplus_{\mathrm{n}+1} h_{y} = \left(h_{a} \oplus_{\mathrm{n}} h_{c}\right)+\left(h_{b} \oplus_{\mathrm{n}} h_{d}\right) \mathbf{i}_{2^{n}+1}.$

    \item The \textbf{multiplication} of $h_x$ and $h_y$ is defined as: $h_{x} \otimes_{\mathrm{n}+1} h_{y} = \left(h_{a} \otimes_{\mathrm{n}} h_{c} \ominus_{\mathrm{n}} \bar{h}_{d} \otimes_{\mathrm{n}} h_{b}\right) +\left(h_{a} \otimes_{\mathrm{n}} h_{d} \oplus_{\mathrm{n}} h_{b} \otimes_{\mathrm{n}} \bar{h}_{c}\right) \mathbf{i}_{2^{n}+1}.$
    When $n \geq 2$, the multiplication is asymmetric.
    
    \item The \textbf{scalar multiplication} of $h_x$ by $\gamma$ is defined as: $\gamma h_x=\gamma h_a + \gamma h_b\mathbf{i}_{2^n+1}.$
    Based on this form of mathematical operations, we can use the low-dimensional algebra operations to study the high-dimensional algebra operations recursively. We will employ these mathematical operations of Cayley-Dickson algebras to design our method HPMRec.

\end{itemize}
\begin{figure*}
    \centering
    \includegraphics[width=1\linewidth]{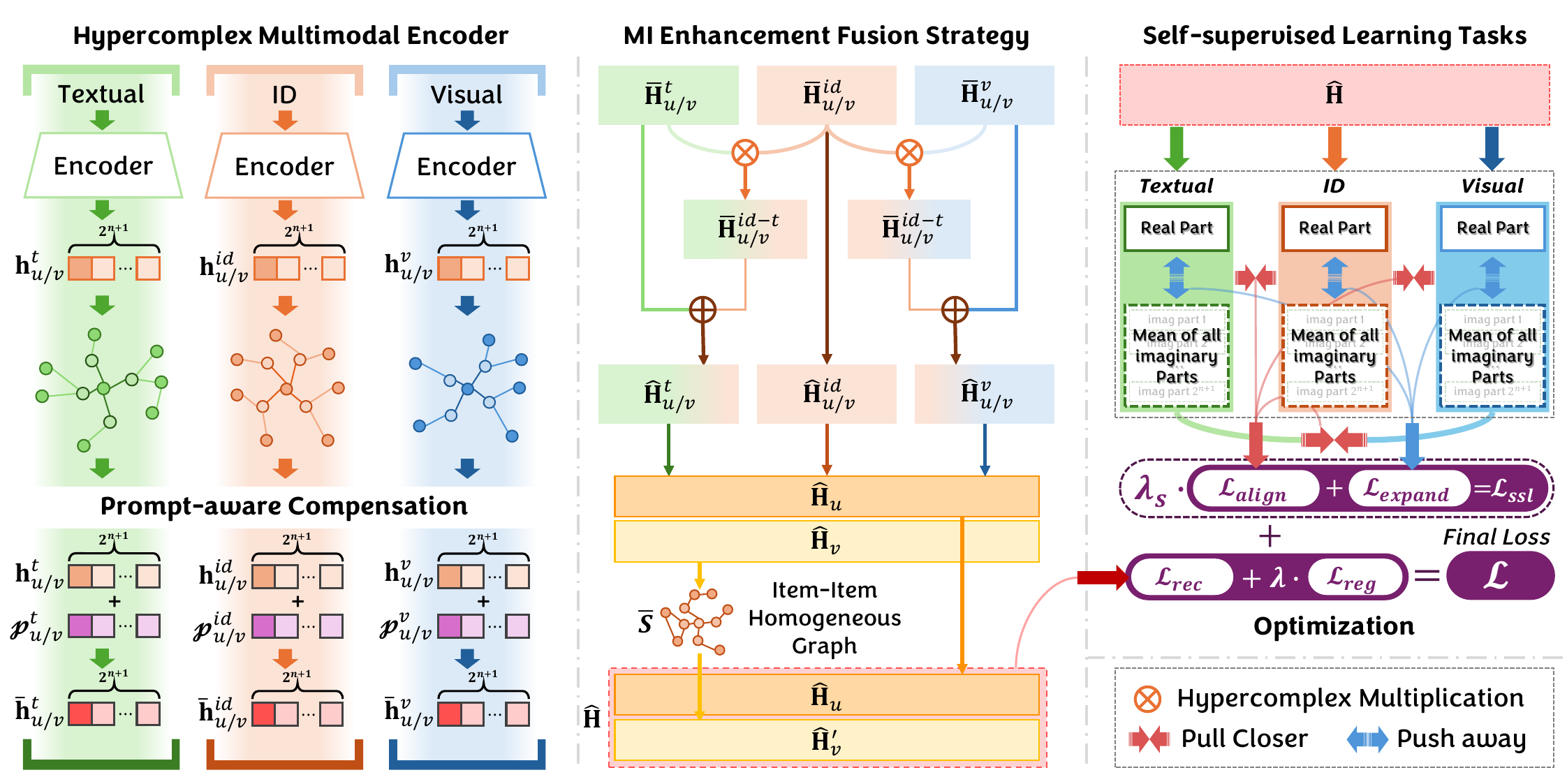}
    \caption{Overall Framework of \model.}
    \label{fig:framework}
\end{figure*}

\section{Methodology}
\label{Sec:methodology}
In this section, we first formulate the problems. Then, we elaborate on the HPMRec framework. Finally, we discuss the optimization process of our HPMRec. The overall framework of HPMRec\footnote{The code is available at: \href{https://github.com/Zheyu-Chen/HPMRec}{https://github.com/Zheyu-Chen/HPMRec}} is shown in Figure~\ref{fig:framework}.
\subsection{Task Definition}
Let $\mathcal{U}=\{u_1,...,u_{|\mathcal{U}|}\}$ denotes user set, $\mathcal{V}=\{v_1,...,v_{|\mathcal{V}|}\}$ denotes item set, and $\mathcal{N}=\mathcal{U} \cup \mathcal{V}$ includes both user and item sets. We conceptualize the user-item graph $\mathcal{G} = (\mathcal{U}, \mathcal{V}, \mathcal{E})$, where $\mathcal{U}, \mathcal{V}$ serve as the graph vertices, and $\mathcal{E}$ denotes the edge set. In the multimodal scenario, each item contains multiple modality features. We introduce modality-specific user/item embedding $\mathbf{h}_{u/v}^m$ for each $u/v$ belonging to the set of modalities $\mathcal{M}$, and we let $\mathbf{H}_{u/v}^m$ denote the entire representation of user/item. Similarly, we let $p_{u/v}^m$ denote the learnable prompt of each user/item, and we let $\mathcal{P}_{u/v}^m$ denote the entire learnable prompt of all users/items. The historical interaction matrix is represented by $\mathcal{R} \in \mathbb{R}^{|\mathcal{U}| \times |\mathcal{V}|}$, in which $r_{uv}=1$ indicates that user $u \in \mathcal{U}$ has engaged with item $v \in \mathcal{V}$ and zero otherwise. 
The goal of our HPMRec is to use the interaction matrix $\mathcal{R}$ and correlation features of each modality $m \in \mathcal{M}$ to predict user $u$'s preference for item $v$ that the user has never engaged with before.

\subsection{Hypercomplex Multimodal Encoder}

To enhance the representation ability, we introduce the hypercomplex algebra as the structure of representation. This structure makes modality-specific features no longer limited to a single vector, which improves the representation diversity of each node during the training process and helps to capture users, items, and their relationships in the user-item interaction graph at a more fine-grained level. Recent works \cite{zhang2021mining, zhou2023tale} find that jointly leveraging user-item heterogeneous graphs and item-item homogeneous graphs can substantially enhance recommendation performance. Building upon these findings, we develop a tailored hypercomplex multimodal encoder architecture to learn modality-specific features through user-item and item-item graphs. 
To be specific, we introduce the Cayley-Dickson algebra, a hypercomplex structure that contains multiple components, as the structure of the representation. Based on this structure, we propose a hypercomplex graph convolution operator that learns node representations, enabling each component to capture diverse modality-specific features.

\subsubsection{Hypercomplex Embedding}
We utilize the Cayley-Dickson construction to encode user and item features with modality $m$ in the hypercomplex space $\mathcal{A}_{n+1}^m$. For user $u$ and item $v$, their hypercomplex embeddings are defined as $\mathbf{h}_u^m$ and $\mathbf{h}_v^m$, respectively. We utilize items' embedding $\mathbf{h}_v^m$ as an example to illustrate this process:
\begin{align}
    \mathbf{h}_v^m=\mathbf{x}_v^m + \mathbf{y}_v^m\mathbf{i}_{2^n+1} = \sum_{k=1}^{2^{n+1}} \mathbf{v}_k^m\mathbf{i}_k,
    \label{eq:9}
\end{align}
where $n \in \mathbb{N}$; $\mathbf{x}_v^m=\sum_{k=1}^{2^n}\mathbf{v}_k^m\mathbf{i}_k \in \mathcal{A}_n^m$ and $\mathbf{y}_v^m=\sum_{k=2^{n}+1}^{2^{n+1}}\mathbf{v}_k^m\mathbf{i}_{k-2^n} \in \mathcal{A}_n^m$ are the subalgebras of $\mathbf{h}_v^m$; $\mathbf{v}_k^m \in \mathbb{R}^{d}$ is the real-valued representation for component $k$, and $d$ denotes the feature dimension. Similar hypercomplex embedding is defined for user $u$.

\subsubsection{Heterogeneous Graph. }
To capture high-order modality-specific features, we construct two \textbf{user-item graphs} $\mathcal{G}$ = $\{\mathcal{G}^{m} \mid m \in \mathcal{M}\}$. Each graph $\mathcal{G}^m$ maintains the same graph structure and only retains the node features associated with each modality. Formally, the message propagation at $l$-th graph convolution layer can be formulated as:

\begin{equation}
    {\mathbf{h}_v^m}{(l)}=\sum_{u \in \mathcal{N}_v} \frac{1}{\sqrt{\left|\mathcal{N}_u\right|} \sqrt{\left|\mathcal{N}_v\right|}} {\mathbf{h}_u^m}{(l-1)}, 
    \label{eq:10}
\end{equation}

\begin{equation}
    \mathbf{h}_u^m{(l)}=\sum_{v \in \mathcal{N}_u} \frac{1}{\sqrt{\left|\mathcal{N}_u\right|} \sqrt{\left|\mathcal{N}_v\right|}} \mathbf{h}_v^m{(l-1)}, 
    \label{eq:11}
\end{equation}
where ${\mathbf{h}_{u/v}^m}{(l)}$ represents the multi-component user/item representation in modality $m$ at $l$-th graph convolution layer. $\mathcal{N}_{u/v}$ denotes the one-hop neighbors of $u/v$ in $\mathcal{G}$. Then, we compute the final user/item embedding of each modality, $\bar{\mathbf{h}}_{u/v}^m$, and describe its aggregation process in detail in Section~\ref{Sec:prompt-aware}.

\subsubsection{Homogeneous Graph. }
We use $k$-NN to establish the \textbf{item-item graph} based on the item features for each modality $m$ to extract significant semantic relations between items. Particularly, we calculate the similarity score $S^m_{v,v^{\prime}}$ between item pair $(v,v^{\prime}) \in \mathcal{V}$ by the cosine similarity $\operatorname{Sim(\cdot)}$ on their modality original features $f_v^m$ and $f_{v^{\prime}}^m$. 

\begin{equation}
S^m_{v,v^{\prime}}=\operatorname{Sim}(f_v^m, f_{v^{\prime}}^m)=\frac{\left(f_v^m\right)^{\top} f_{v^{\prime}}^m}{\left\|f_v^m\right\|\left\| f_{v^{\prime}}^m\right\|}.
\label{eq:12}
\end{equation}

We only retain the top-$k$ neighbors:

\begin{equation}
\bar{S}^m_{v,v^{\prime}} = \begin{cases} S^m_{v,v^{\prime}} & \text { if } S^m_{v,v^{\prime}} \in \text { top-} k (S^m_{v,p} \mid p \in \mathcal{V}) \\ 0 & \text { otherwise }\end{cases},
\label{eq:13}
\end{equation}
where $\bar{S}^m_{v,v^{\prime}}$ represents the edge weight between item $v$ and item $v^{\prime}$ within modality $m$. Thereafter, we further build a unified item-item graph $\bar{S}$ by aggregating all modality-specified graphs $\bar{S}^m$:

\begin{equation}
\bar{S} = \sum_{m \in \mathcal{M}} \alpha^m \bar{S}^m.
\label{eq:14}
\end{equation}

Inspired by \cite{zhou2023tale}, we freeze each item-item graph after initialization to eliminate the computational cost of the item-item graph during training. In addition, $\alpha_m$ is a trainable parameter with the same initial value for each modality.

\subsection{Prompt-aware Compensation}
\label{Sec:prompt-aware}
Due to the multi-component structure of hypercomplex embedding, each component of each modality is able to learn different features, which poses a challenge to the efficient use of these representations. As these highly diverse components deviate from the initial semantic space, the representations of multiple components are misaligned. Directly concatenating\footnote{If we accumulate or calculate the mean of the representations of these components, we can avoid this problem, but the diverse representations learned in each component will be lost. It goes against the purpose of adopting the multi-component structure.} these components will result in a low-quality user/item representation and will even lose the core modality-specific features.

To this end, we designed the learnable prompt $p \in \mathbb{R}^{d \cdot 2^{n+1}}$ to independently compensate the features learned by each component. The final embedding for each modality $m$ is as follows: 
\begin{equation}
    \bar{\mathbf{h}}_{u/v}^m = \sum_{l=0}^L\mathbf{h}_{u/v}^ m{(l)} + p_{u/v}^m, 
    \label{eq:15}
\end{equation}
where $L$ is the number of user-item graph layers. In addition, in the message passing process of GCN, user/item representation will inevitably tend to be the same as their neighbors'. HPMRec allows representations to learn diversity and then uses the learnable prompt for dynamic compensation, which fundamentally alleviates the over-smoothing problem. Specifically, the learnable prompt keeps the core modality-specific features in each component, and the diversity is retained. Therefore, the diversity ensures that the representations are not over-smoothing.

\subsection{MI Enhancement Fusion Strategy}
Previous works \cite{liu2021interest, mao2021ultragcn} adopt linear strategies to fuse modalities, such as weighted sums or concatenations. However, linear fusion can not sufficiently mine the latent relation among modalities. 

Therefore, we apply hypercomplex algebraic multiplication to naturally build nonlinear relations among different modalities' components, enhancing the representation's ability to mine latent cross-modality features. 
\begin{align}
    \bar{\mathbf{H}}_{u/v}^{id-v} = \bar{\mathbf{H}}_{u/v}^{id} \otimes_{n+1} \bar{\mathbf{H}}_{u/v}^{v},
    \label{eq:16}
\end{align}
\begin{align}
    \bar{\mathbf{H}}_{u/v}^{id-t} = \bar{\mathbf{H}}_{u/v}^{id} \otimes_{n+1} \bar{\mathbf{H}}_{u/v}^{t}.
    \label{eq:17}
\end{align}

When two hypercomplex algebras are multiplied, the product incorporates the nonlinearities and higher-order dependencies between the original algebras \cite{rosas2019quantifying}. Next, we add it back to the original modalities to enhance their cross-modality features, which is beneficial to modality fusion. 
\begin{align}
    \hat{\mathbf{H}}_{u/v}^{v} = \bar{\mathbf{H}}_{u/v}^{v} + \epsilon_1 \cdot \bar{\mathbf{H}}_{u/v}^{id-v},    
    \label{eq:18}
\end{align}
\begin{align}
    \hat{\mathbf{H}}_{u/v}^{t} = \bar{\mathbf{H}}_{u/v}^{t} + \epsilon_2 \cdot \bar{\mathbf{H}}_{u/v}^{id-t}, 
    \label{eq:19}
\end{align}
where $\epsilon_1, \epsilon_2$ are trainable parameters to control the MI enhancement strength, which are empirically initialized with 0.1. To simplify the formula expression, we let $\hat{\mathbf{H}}_{u/v}^{id} = \bar{\mathbf{H}}_{u/v}^{id}$. Then we calculate the final user/item representations:
\begin{equation}
    \hat{\mathbf{H}}_{u/v} = \operatorname{Con}(\beta^m \hat{\mathbf{H}}_{u/v}^{m} \mid m \in \mathcal{M}),
    \label{eq:20}
\end{equation}
where the attention weight $\beta^m$ is a trainable parameter, which is initialized with equal value for each modality. Then we enhance the item representations $\hat{\mathbf{H}}_{v}$ by item-item graphs $\bar{S}$. Ultimately, we fuse the final user representation and enhanced item representations to get the final representation:
\begin{equation}
   \hat{\mathbf{H}}=\operatorname{Con}(\hat{\mathbf{H}}_{u}, \hat{\mathbf{H}}_{v}^{'}), \quad\hat{\mathbf{H}}_{v}^{'} = \hat{\mathbf{H}}_{v} + {\bar{S}} \cdot \hat{\mathbf{H}}_{v}.
   \label{eq:21}
\end{equation}

\subsection{Self-Superised Learning Tasks}
\label{Sec:self-supervised learning}

\subsubsection{Cross-modality Alignment}
We employ self-supervised learning, taking the mean of the Manhattan distance\footnote{We also considered adopting the Euclidean distance, but since the performance difference was almost the same and the performance consumption was higher, we chose an easy-to-use and effective method. In addition, compared with high-order distances, low-order metrics are more stable and less susceptible to extreme values, which is conducive to the stability of model training.} to align ID-visual, ID-textual, and visual-textual modality pairs. Formally:
\begin{equation}
\mathcal{L}_{align} = -\frac{1}{|\mathcal{N}|}\sum_{n \in \mathcal{N}}\sum_{(a,b) \in \mathcal{C}} \left[|\hat{\mathbf{h}}^a_n - \hat{\mathbf{h}}^b_n|\right],
\label{eq:22}
\end{equation}
where $\mathcal{C} \in \{(id,v),(id,t),(v,t)\}$ denote set of modality pairs. This task brings the representations of each modality closer, which is beneficial to modality fusion and final rating prediction.

\subsubsection{Real-Imag Discrepancy Expansion}
We expand the discrepancy among different components to enhance the diversity of user/item representation. Specifically, we directly take the mean of the Manhattan distance between the real part and the mean of all imaginary parts of each modality.
\begin{equation}
\mathcal{L}_{expand} = -\frac{1}{|\mathcal{N}|}\sum_{n \in \mathcal{N}}\sum_{m \in \mathcal{M}} \left[|\hat{\mathbf{c}}^m_n - \mathbb{E}[\hat{\mathbf{d}}^m_n]|\right],
\label{eq:23}
\end{equation}
where $\mathcal{M} \in \{id, v, t\}$, $\mathcal{N} = \mathcal{U} \cup \mathcal{V}$, and $\mathbb{E}[\cdot]$ represents the mean calculation of the component-level. $\hat{\mathbf{c}}^m_n$ and $\hat{\mathbf{d}}^m_n$ denote the real part and imaginary parts of node $n$'s representation $\hat{\mathbf{h}}^m_n$, respectively.

Here is the final self-supervised learning loss, formally:
\begin{align}
    \mathcal{L}_{ssl} = \mathcal{L}_{align} + \mathcal{L}_{expand}.
    \label{eq:24}
\end{align}

\subsection{Optimization}
We adopt LightGCN \cite{he2020lightgcn} as the backbone model and employ the Bayesian Personalized Ranking (BPR) loss \cite{rendle2009bpr} as the primary optimization objective. The BPR loss is specifically designed to improve the predicted preference distinction between positive and negative items for each triplet $(u, p, n) \in \mathcal{D}$, where $\mathcal{D}$ represents the training dataset. In this context, the positive item $p$ is one with which user $u$ has interacted, while the negative item $n$ is randomly selected from the set of items that user $u$ has not interacted with. Formally:

\begin{equation} 
    \mathcal{L}_{rec} = \sum_{(u, p, n) \in \mathcal{D}} - \log(\sigma(y_{u,p} - y_{u,n})) + \lambda \cdot \|\mathbf{\Theta}\|^2_2, 
    \label{eq:25}
\end{equation}
where $\sigma$ represents the sigmoid function, and $\lambda$ controls the strength of $L_2$ regularization, and $\mathbf{\Theta}$ denotes the parameters subject to regularization. The terms $y_{u,p}$ and $y_{u,n}$ correspond to the ratings of user $u$ for the positive item $p$ and the negative item $n$, respectively, computed as $\hat{\mathbf{h}}_u^\top \cdot \hat{\mathbf{h}}_p$ and $\hat{\mathbf{h}}_u^\top \cdot \hat{\mathbf{h}}_n$. The final loss function is given by:
\begin{equation}
\mathcal{L} = \mathcal{L}_{rec} + \lambda_s \mathcal{L}_{ssl},
\label{eq:26}
\end{equation}
where $\lambda_{s}$ is the self-superised learning balancing hyper-parameter.

To provide a clearer overview of our HPMRec, we summarize the learning process of HPMRec in Algorithm~\ref{al}.

\begin{algorithm}[t]
\caption{Learning Process of HPMRec}
\label{al}
\begin{algorithmic} [1] 
\STATE \textbf{Input:} $\mathcal{U}$, $\mathcal{V}$, $\mathcal{M}$, $\mathcal{G}$, node set $\mathcal{N} = \mathcal{U} \cup \mathcal{V}$, layer number $L$ of heterogeneous graph $\mathcal{G}$.
\STATE \textbf{Output:} Optimization loss $\mathcal{L}$
\STATE Initialize $\mathbf{H}_u^m$, $\mathbf{H}_v^m$, $\mathcal{P}_u^m$, $\mathcal{P}_v^m$;
\FOR{$l = 1...L$}
    \STATE Conduct message passing in the heterogeneous graph $\mathbf{h}_v^m(l)$ $\gets$ $\mathbf{h}_u^m(l-1)$ with Eq.\ref{eq:10}, or $\mathbf{h}_u^m(l)$ $\gets$ $\mathbf{h}_v^m(l-1)$ with Eq.\ref{eq:11};
\ENDFOR
\STATE Get pormpt-aware compensated embedding $\bar{\mathbf{h}}_u^m$, $\bar{\mathbf{h}}_v^m$ for each modality with Eq.\ref{eq:15};
\STATE Construct the unified item-item graph $\bar{S}$ with Eq.\ref{eq:12}-\ref{eq:14};
\STATE Represent all node embeddings $\mathbf{h}$ as the entire node representation $\mathbf{H}$.
\STATE Apply hypercomplex multiplication with Eq.\ref{eq:16}-\ref{eq:17};
\STATE Get enhanced representation $\hat{\mathbf{H}}^v_{u/v}$ and $\hat{\mathbf{H}}^t_{u/v}$ with Eq.\ref{eq:18}-\ref{eq:19};
\STATE Attentively fuse all modality representations $\hat{\mathbf{H}}_{u/v}$ $\gets$ $\hat{\mathbf{H}}^m_{u/v}$ with Eq.\ref{eq:20};
\STATE Get final representation $\hat{\mathbf{H}}$ by item-item graphs $\bar{S}$ with Eq.\ref{eq:21}.
\STATE Calculate self-supervised learning loss $\mathcal{L}_{ssl}$ with Eq.\ref{eq:22}-\ref{eq:24};
\STATE Calculate adaptive BPR loss $\mathcal{L}_{rec}$ with Eq.\ref{eq:25};
\STATE Get final optimization loss $\mathcal{L}$ with Eq.\ref{eq:26}.
\end{algorithmic}
\end{algorithm}

\section{Experiments}
In this section, we conduct comprehensive experiments to evaluate the performance of our \model\space framework on four widely used real-world datasets. The following five research questions can be well answered through experimental results:
\textbf{RQ1: }Does HPMRec outperform the state-of-the-art conventional and multimodal recommendation methods?
\textbf{RQ2: }What impact do the key modules of our HPMRec framework have on its overall performance?
\textbf{RQ3: }How does the representation scaling strategy affect the performance-efficiency trade-off?
\textbf{RQ4: }How efficient is HPMRec compared with various state-of-the-art recommender systems?
\textbf{RQ5: }How do different hyper-parameter settings impact the overall performance of HPMRec?

\subsection{Datasets and Evaluation Metrics}
To evaluate the performance of our proposed \model\space in the recommendation task, we perform comprehensive experiments on four widely used Amazon datasets \cite{mcauley2015image}: Office, Baby, Sports, and Clothing. These datasets offer both product descriptions and images. In line with previous works \cite{xu2025survey, xu2025mentor}, we preprocess the raw data with a 5-core setting for both items and users. Additionally, we utilize pre-extracted 4096-dimensional visual features and obtain 384-dimensional textual features using a pre-trained sentence transformer \cite{zhou2023mmrecsm}. For a fair evaluation, we employ two widely recognized metrics: Recall@$K$ (R@$K$) and NDCG@$K$ (N@$K$). We present the average metrics for all users in the test dataset for both $K$ = 10 and $K$ = 20. We adhere to the standard procedure \cite{zhou2023tale} with a random data split of 8:1:1 for training, validation, and testing.

\begin{table}[h]
    \centering
\caption{Statistics of datasets.}
\label{tab:dataset_statistics}
    \begin{tabular}{ccccc}
         \toprule
         \textbf{Datasets}&  \textbf{\#Users}&  \textbf{\#Items}&  
         \textbf{\#Interactions}& \textbf{Sparsity}\\ 
         \midrule
         \textbf{Office} & 4,905 & 2,420 & 53,258 &99.55\%\\
         \textbf{Baby} & 19,445 & 7,050 & 160,792 & 99.88\%\\
         \textbf{Sports} & 35,598 & 18,357 & 296,337 & 99.95\%\\
         \textbf{Clothing}&  39,387&  23,033&  278,677& 99.97\%\\
         \bottomrule
    \end{tabular}
\end{table}

\begin{table*}[!ht]
\centering
\small
\tabcolsep=0.1cm
\caption{Performance comparison of baselines and \model(our) in terms of Recall@K(R@K) and NDCG@K(N@K).}
\label{tab:comparison results}
\begin{tabular}{rllllcccccccccccc}
\toprule

\multirow{2.5}{*}{\textbf{Model }}& \multicolumn{4}{c}{\textbf{Office}}&\multicolumn{4}{c}{\textbf{Baby}} & \multicolumn{4}{c}{\textbf{Sports}} & \multicolumn{4}{c}{\textbf{Clothing}}\\ \cmidrule(lr){2-5} \cmidrule(lr){6-9} \cmidrule(lr){10-13} \cmidrule(lr){14-17}& R@10   & {R@20}   & N@10   & N@20    
&R@10   & {R@20}   & N@10   & {N@20}   & R@10   & {R@20}   & N@10   & N@20    & R@10   & {R@20}   & N@10   &N@20    \\
\midrule
 MF-BPR &  0.0572& 0.0951 & 0.0331& 0.0456
&0.0357& 0.0575& 0.0192& 0.0249
& 0.0432& 0.0653& 0.0241&0.0298
 & 0.0187& 0.0279& 0.0103&0.0126\\
 LightGCN &  0.0791& 0.1189& 0.0459& 0.0583
&0.0479& 0.0754& 0.0257& 0.0328
& 0.0569& 0.0864& 0.0311&0.0387
 & 0.0340& 0.0526& 0.0188&0.0236\\
 SimGCL &  0.0799& 0.1239& 0.0470& 0.0595
&0.0513 & 0.0804 & 0.0273 & 0.0350 
& 0.0601 & 0.0919 & 0.0327 &0.0414 
 & 0.0356& 0.0549& 0.0195&0.0244\\
 LayerGCN &  0.0825& 0.1213& 0.0486& 0.0593
&0.0529& 0.0820& 0.0281& 0.0355& 0.0594& 0.0916& 0.0323&0.0406 & 0.0371& 0.0566& 0.0200&0.0247\\ 
\hline

{VBPR}& 0.0692& 0.1084& 0.0422& 0.0531
&0.0423& 0.0663& 0.0223& 0.0284
& 0.0558& 0.0856& 0.0307& 0.0384
 & 0.0281& 0.0415& 0.0158&0.0192\\
{MMGCN }& 0.0558& 0.0926& 0.0312& 0.0413
&0.0378& 0.0615& 0.0200& 0.0261
& 0.0370& 0.0605& 0.0193& 0.0254
 & 0.0218& 0.0345& 0.0110&0.0142\\
{DualGNN}& 0.0887& 0.1350& 0.0505& 0.0631
&0.0448& 0.0716& 0.0240& 0.0309
& 0.0568& 0.0859& 0.0310& 0.0385
 & 0.0454& 0.0683& 0.0241&0.0299\\
{LATTICE}& 0.0969& 0.1421& \underline{0.0562}& \underline{0.0686}
&0.0547& 0.0850& 0.0292& 0.0370
& 0.0620& 0.0953& 0.0335& 0.0421
 & 0.0492& 0.0733& 0.0268&0.0330\\
{FREEDOM}& \underline{0.0974}& \underline{0.1445}& 0.0549& 0.0669
&0.0627& \underline{0.0992}& 0.0330& 0.0424
& 0.0717& \underline{0.1089}& 0.0385& \underline{0.0481}
 & \underline{0.0629}& \underline{0.0941}& \underline{0.0341}&\underline{0.0420}\\
{SLMRec}& 0.0790& 0.1252& 0.0475& 0.0599
&0.0529& 0.0775& 0.0290& 0.0353
& 0.0663& 0.0990& 0.0365&  0.0450
 & 0.0452& 0.0675& 0.0247&0.0303\\ 
{BM3}& 0.0715& 0.1155& 0.0415& 0.0533
&0.0564& 0.0883& 0.0301& 0.0383
& 0.0656& 0.0980& 0.0355& 0.0438
 & 0.0422& 0.0621& 0.0231&0.0281\\
{MMSSL}& 0.0794& 0.1273& 0.0481& 0.0610
&0.0613& 0.0971& 0.0326& 0.0420
& 0.0673& 0.1013& 0.0380& 0.0474
 & 0.0531& 0.0797& 0.0291&0.0359\\
{LGMRec}& 0.0959& 0.1402& 0.0514& 0.0663
&\underline{0.0639}& 0.0989& \underline{0.0337}& \underline{0.0430}
& \underline{0.0719}& 0.1068& \underline{0.0387}& 0.0477
 & 0.0555& 0.0828& 0.0302&0.0371\\
{DiffMM}& 0.0733& 0.1183& 0.0439& 0.0560
&0.0623& 0.0975& 0.0328& 0.0411& 0.0671& 0.1017& 0.0377& 0.0458 & 0.0522& 0.0791& 0.0288&0.0354\\ \midrule
\textbf{\model\space}& \textbf{0.1092}& \textbf{0.1632}& \textbf{0.0632}& \textbf{0.0778}&\textbf{0.0667}& \textbf{0.1033}& \textbf{0.0357}&  \textbf{0.0451}& \textbf{0.0751}& \textbf{0.1129}& \textbf{0.0410}& \textbf{0.0507}& \textbf{0.0658}& \textbf{0.0963}& \textbf{0.0351}& \textbf{0.0429}\\
\bottomrule
\end{tabular}
\end{table*}

\subsection{Baselines}
To comprehensively evaluate the effectiveness of \model, we conduct a systematic comparison with state-of-the-art methods, categorized into traditional recommendation methods (focusing on collaborative filtering and graph-based learning) and multimodal recommendation methods (leveraging multiple modalities such as visual and textual features). Below, we provide a concise yet informative overview of each baseline method.

\noindent1) Conventional recommendation methods:
\begin{itemize}[leftmargin=*]
 \item \textbf{MF-BPR} \cite{rendle2012bpr}: optimized with Bayesian Personalized Ranking (BPR) loss, designed for learning user and item embeddings from implicit feedback.
 \item \textbf{LightGCN} \cite{he2020lightgcn}: removes unnecessary modules: nonlinear activations to improve recommendation performance.
 \item \textbf{SimGCL} \cite{yu2022graph}: enhances representation robustness by injecting controlled noise into embeddings.
 \item \textbf{LayerGCN} \cite{zhou2023layer}: alleviating LightGCN's over-smoothing issue via residual connections, refining layer-wise aggregation for deeper GCNs.
 \end{itemize}

\noindent2) Multimodal recommendation methods:
\begin{itemize}[leftmargin=*]
 \item \textbf{VBPR} \cite{he2016vbpr}: extends matrix factorization by incorporating visual and textual features as side information for items.
 \item \textbf{MMGCN} \cite{wei2019mmgcn}: employs separate GCNs per modality and fuses modality-specific predictions for final recommendations.
 \item \textbf{DualGNN} \cite{wang2021dualgnn}: introduces a user-user graph to model latent preference patterns beyond user-item interactions.
 \item \textbf{LATTICE} \cite{zhang2021mining}: constructs an item-item graph to capture high-order semantic relationships among items.
 \item \textbf{FREEDOM} \cite{zhou2023tale}: enhances LATTICE by freezing the item-item graph and denoising the user-item graph.
 \item \textbf{SLMRec} \cite{tao2022self}: employs node self-discrimination to uncover multimodal item patterns.
 \item \textbf{BM3} \cite{zhou2023bootstrap}: simplifies self-supervised learning via dropout-based representation perturbation.
 \item \textbf{MMSSL} \cite{wei2023multi}: combines modality-aware adversarial training with cross-modal contrastive learning to disentangle shared and modality-specific features.
 \item \textbf{LGMRec} \cite{guo2024lgmrec}: unifies local (graph-based) and global (hypergraph-based) embeddings for multimodal recommendation.
 \item \textbf{DiffMM} \cite{jiang2024diffmm}: leverages modality-aware graph diffusion to improve user representation learning.
 \end{itemize}

\subsection{Experimental Settings}
Following the basic settings of previous works \cite{zhou2023tale}, we implement HPMRec in PyTorch and optimize with the Adam optimizer \cite{kingma2014adam}. We apply Xavier initialization \cite{glorot2010understanding} for all initial random embeddings. As for hyper-parameter settings on HPMRec, we perform a grid search on the user-item heterogeneous graph $\mathcal{G}$'s GCN layer number $L$  in \{1, 2, 3\}, regularization balancing hyper-parameter $\lambda$ in \{1$e^{-2}$, 1$e^{-3}$, 1$e^{-4}$\}, self-supervised learning balancing hyper-parameter $\lambda_{s}$ in \{1$e^{-2}$, 1$e^{-3}$, 1$e^{-4}$\}. We set $n$ in hypercomplex algebra to \{0, 1, 2, 3\}, which indicate the components number $2^{n+1}$ in \{2, 4, 8, 16\}. We fix the learning rate as 1$e^{-4}$, and adopt a single-layer GCN in the item-item homogeneous graph. The $k$ of top-$k$ in the item-item graph is set as 10. For convergence consideration, we fixed the early stopping at 20. Following the settings of \cite{zhou2023mmrecsm}, we update the best record by utilizing Recall@20 on the validation dataset as the indicator. All the experiments were conducted on the NVIDIA GeForce RTX 3090 GPU.

\subsection{Overall Performance (RQ1)}
Detailed experiment results are shown in Table~\ref{tab:comparison results}. The optimal results are highlighted in bold, while the suboptimal ones are underlined. We have the following key observations:

\textbf{Our framework consistently outperforms all baselines across all datasets and evaluation metrics,} demonstrating both its effectiveness across datasets with varying scales and sparsity.

\textbf{The multi-component structure of hypercomplex embeddings and the prompt-aware compensation mechanism effectively enhance the ability of representation.} The hypercomplex embedding provides multiple components to capture the diverse modality-specific features, and the learnable prompt is able to dynamically compensate for the misalignment of multiple components and the loss of core modality-specific features. And due to the diversity of components, the user/item representation is not the same as the neighbors', so that it keeps the representation away from over-smoothing problems. Compared to previous works \cite{zhou2023layer, mao2021ultragcn, liu2021interest} that utilize static optimization to alleviate the over-smoothing problem, we have superior performance.

\textbf{Our MI enhancement fusion strategy and self-supervised learning tasks also positively impact overall performance, making the framework more robust.} Notably, our nonlinear fusion strategy outperforms existing linear and attention-based strategies through deeper latent relation exploration. Comprehensive ablation studies in Section~\ref{Sec:ablation study} will systematically dissect each module's contribution to overall performance.

\begin{table}[!h]
    \centering
    \small
    \caption{Performance Comparison on variants of HPMRec.}
\label{tab:ablation study}
\begin{tabular}{ccccc}
\toprule
\multirow{2.5}{*}{\textbf{Variant}}                                 &\multicolumn{2}{c}{\textbf{Baby}}&\multicolumn{2}{c}{\textbf{Sports}} \\ \cmidrule(lr){2-3} \cmidrule(lr){4-5}& Recall@10   & NDCG@10   & Recall@10   & NDCG@10    \\
\midrule
{$w/o$ Prompt}                    &  0.0592&  0.0312& 0.0541& 0.0297\\
{$w/o$ MI}                    & 0.0657& 0.0351& 0.0739& 0.0403\\
{$w/o$ SSL}                & 0.0651& 0.0343& 0.0742& 0.0401\\
{$w/o$ Enhance}                & 0.0640& 0.0346& 0.0733& 0.0398\\
{$w$ Explicit}                & 0.0654& 0.0348& 0.0736& 0.0403\\
\midrule
{HPMRec-Split}                & 0.0654& 0.0354& 0.0708& 0.0381\\
{HPMRec-MLP}                & 0.0582& 0.0319& 0.0682& 0.0369\\
\midrule
{HPMRec}                & \textbf{0.0667}& \textbf{0.0357}& \textbf{0.0751}& \textbf{0.0410}\\
\bottomrule
\end{tabular}
\end{table}

\begin{figure}[h]
    \centering
    \includegraphics[width=1\linewidth]{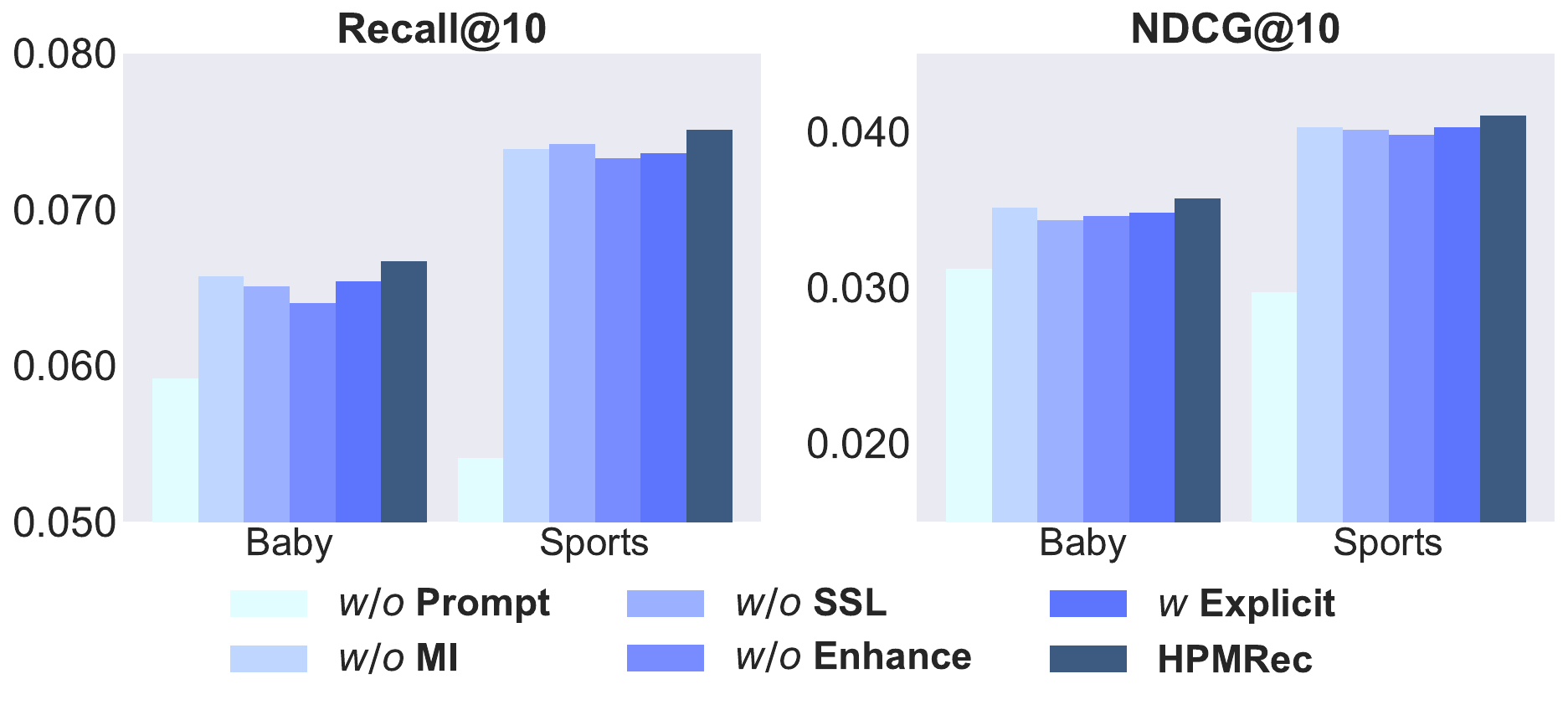}
    \caption{Effect of key modules in HPMRec.}
    \label{fig:bar1}
\end{figure}

\subsection{Ablation Study (RQ2 \& RQ3)}
\label{Sec:ablation study}
In this section, we conduct extensive experiments to evaluate the effectiveness of each module in HPMRec. We also explored how the node representation's feature dimension scaling strategy affects the performance-efficiency trade-off in resource-constrained scenarios.
\subsubsection{Effectiveness of key modules of HPMRec (RQ2)}
\begin{itemize}[leftmargin=*]
    \item \textbf{HPMRec $w/o$ Prompt: }Remove the learnable prompt from each node representation.
    \item \textbf{HPMRec $w/o$ MI: }Remove the MI enhancement operation from the fusion stage.
    \item \textbf{HPMRec $w/o$ SSL: }Remove the self-supervised learning tasks.
    \item \textbf{HPMRec $w/o$ Enhance: }Remove both the MI enhancement operation and the self-supervised learning tasks.
    \item \textbf{HPMRec $w$ Explicit: }Explicitly align the prompt with the initial layer node representation.
\end{itemize}

Detailed ablation study experiment results are shown in Table~\ref{tab:comparison results} and Figure~\ref{fig:bar1}. We have the following key observations:

Our multi-component hypercomplex embedding comprehensively explores the representational potential of each node, enabling rich and detailed feature extraction. However, introducing hypercomplex embeddings inevitably results in component-level misalignment, hindering effective representation learning. To this end, we propose a prompt-aware compensation mechanism that adaptively aligns the semantic spaces of different components. The performance of variant \textbf{HPMRec $w/o$ Prompt} shows that the prompt-aware compensation mechanism is significant for our framework, and adopting hypercomplex embedding with multi-component structure alone is not feasible in multimodal scenarios. According to the result of the variant \textbf{HPMRec $w/o$ Enhance}, with only the hypercomplex embedding and learnable prompt, we still surpass all baselines, indicating the significant effectiveness of these two modules.

The performance degradation observed in variant \textbf{HPMRec $w/o$ MI} and variant \textbf{HPMRec $w/o$ SSL} confirms the effectiveness of our MI enhancement fusion strategy and self-supervised learning tasks. These modules contribute not only to performance improvements but also to better framework robustness. In particular, the self-supervised learning module facilitates cross-modal alignment among ID, visual, and textual representations and enhances the diversity of different components for hypercomplex embedding. Moreover, our MI enhancement fusion strategy, which is based on hypercomplex multiplication, a naturally nonlinear calculation, outperforms existing linear and attention-based fusion strategies, demonstrating its superior capability in capturing the cross-modality features.

In the variant \textbf{HPMRec $w$ Explicit}, we design the closest explicit guidance to the motivation of designing the learnable prompt, that is, aligning the learnable prompt with initial layer node representations. However, the performance degradation observed in variant \textbf{HPMRec $w$ Explicit} demonstrates that the explicit guidance will harm the ability of the learnable prompt. Therefore, we employ no optimization task for the learnable prompt explicitly, only utilize the main recommendation task and self-supervised learning tasks to implicitly benefit its dynamic optimization. Our HPMRec's higher performance results shows the learnable prompt can perform better in implicit guidance than explicit guidance. We will further discuss how to maximize the ability of prompt in Section~\ref{Sec:maximize}.

\begin{figure}[h]
    \centering
    \includegraphics[width=1\linewidth]{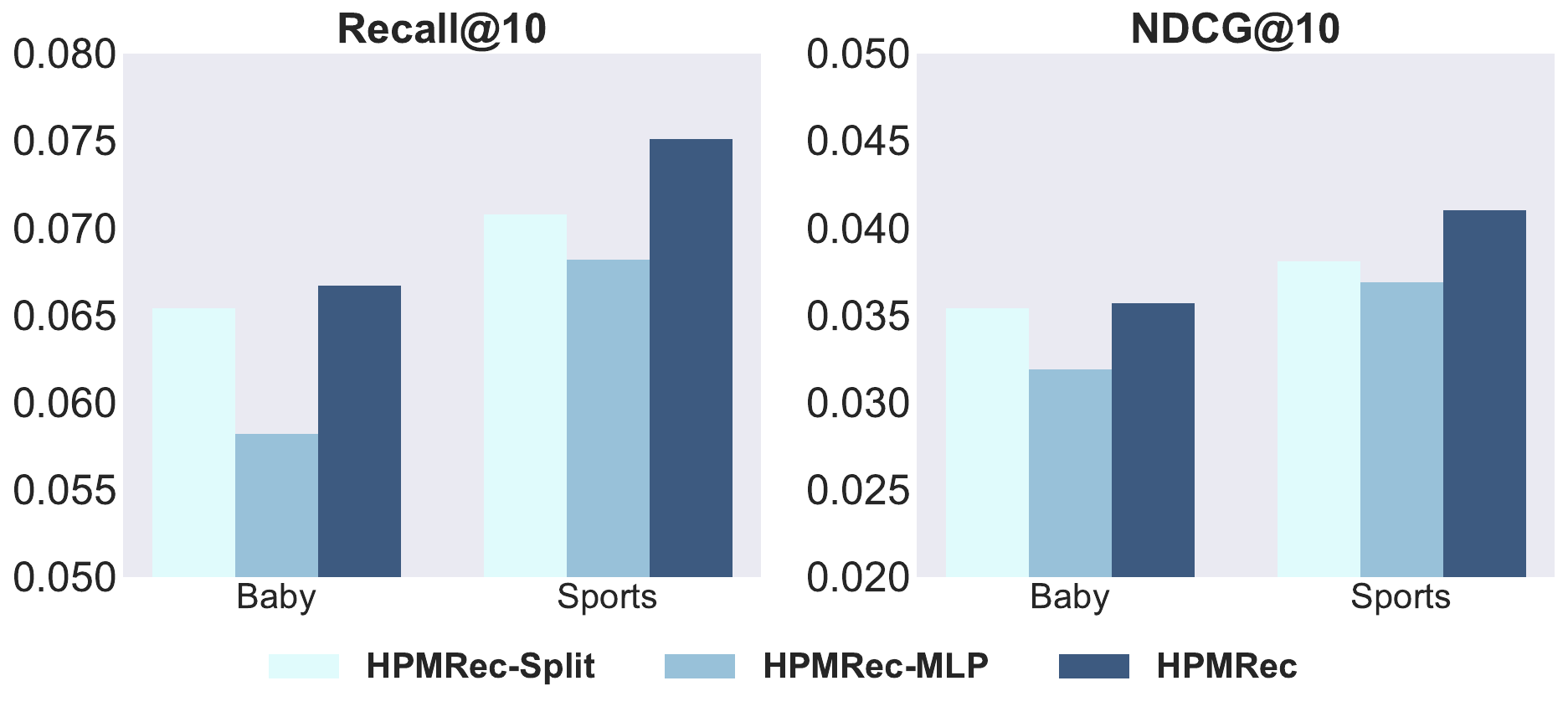}
    \caption{Effect of Dimension Scale Trade-off Strategy.}
    \label{fig:bar2}
\end{figure}

\subsubsection{Feature dimension scale trade-off strategy (RQ3)}

In Section~\ref{Sec:hyperparameter}, we found that the best performance of the framework does not gain advantages when the component number $2^{n+1}$ is very large, which makes it unnecessary to consider the computational resource consumption caused by the limited representation's feature dimension increasing, and we also conduct effieiency study in Section~\ref{Sec:efficiency} to demonstrate the competitive efficiency of our framework. However, to consider a more comprehensive computation resources scenario, we design the following variant to explore the consumption-performance trade-off. Detailed ablation study experiment results are shown in Table~\ref{tab:comparison results} and Figure~\ref{fig:bar2}.
\begin{itemize}[leftmargin=*]
    \item \textbf{HPMRec-Split: }In the modality information encoding stage, each modality representation is partitioned into $2^{n+1}$ equal segments as components.
    \item \textbf{HPMRec-MLP: }In the modality information encoding stage, this variant employs an MLP to compress the feature dimension of each component from $d$ to $d /2^ {n+1}$, which will keep the feature dimension of each node representation to $d$, instand of $d\cdot2^ {n+1}$.
\end{itemize}

The dimensionality reduction of variant \textbf{HPMRec-Split} inevitably sacrifices some representation capacity, thereby capping the framework's potential performance. Furthermore, the simple dimension compression of variant \textbf{HPMRec-MLP} makes each component small and similar, meanwhile loses the diversity of user/item representation, which leads to a significant performance degradation.

The performance results of the two variants in Table~\ref{tab:ablation study} show that a sufficient feature dimension is crucial for exploring multimodal information. If the feature dimension scale is simply limited, the representation's capability will be reduced due to the lack of diverse modality-specific features. In addition, we found that variant \textbf{HPMRec-Split} has higher performance than variant \textbf{HPMRec-MLP}, which shows that although the representation is divided into multiple components, the origin modality-specific features are complete, ensuring a certain richness and diversity. It can still restore some representation capabilities under the compensation of the learnable prompt. The variant \textbf{HPMRec-MLP}, which directly compresses the feature dimension of the representation to a very small scale, not only fails to retain the core modality-specific features but also loses diversity. Although prompt has the ability to compensate the core modality-specific features, the user/item representation's diversity has been lost. This situation will be more obvious as the hypercomplex dimension grows, because the feature dimension of its single component will shrink as the component number grows. 

In summary, variant \textbf{HPMRec-Split} reduces both computation and memory requirements in resource-constrained scenarios, which is suitable for scenarios with extremely limited computational resources, whereas variant \textbf{HPMRec-MLP} fails to achieve a favorable trade-off between efficiency and effectiveness due to the loss of component diversity. This ablation study experiment proves that multimodal information requires a sufficient feature dimension scale to explore diverse features, which is consistent with our motivation for using hypercomplex embedding. And the higher performance of variant \textbf{HPMRec-Split} also indirectly proves the effectiveness of our prompt-aware compensation mechanism.

\subsection{Efficiency Study (RQ4)}
\label{Sec:efficiency}
We report the training time per epoch and memory usage of HPMRec and baselines in Table~\ref{tab:efficiency}\footnote{The results are under the best hyper-parameter settings on each dataset. }. After analyzing the results of efficiency, we found that our framework maintains competitive efficiency in terms of training time per epoch and memory usage. 

Thanks to the multi-component structure, the node representation contains rich and diverse modality-specific features. Thus, these powerful representations enable the framework to achieve the best performance with fewer convolutional layers on large but sparse datasets (e.g., Clothing), which means that we are not constrained by the high resource consumption of GCN, and have stable training time on all datasets.

\begin{table}[h]
\small
\caption{Comparison of our HPMRec against state-of-the-art baselines on efficiency. (Time: s/Epoch; Memory: GB)}
\label{tab:efficiency}
\begin{tabular}{ccccccc}
\toprule
\multicolumn{1}{c}{Dataset} & \multicolumn{2}{c}{\textbf{Baby}} & \multicolumn{2}{c}{\textbf{Sports}} & \multicolumn{2}{c}{\textbf{Clothing}} \\
\midrule
\multicolumn{1}{c}{Metrics} & \multicolumn{1}{c}{Time} & \multicolumn{1}{c}{Memory} & \multicolumn{1}{c}{Time} & \multicolumn{1}{c}{Memory} & \multicolumn{1}{c}{Time} & \multicolumn{1}{c}{Memory} \\
\midrule
DualGNN & 5.63 & 2.05 & 11.59 & 2.81 & 14.19 & 3.02 \\
MMGCN & 4.09 & 2.69 & 14.93 & 3.91 & 17.48 & 4.24 \\
LATTICE & 3.20 & 4.53 & 11.07 & 19.93 & 16.53 & 28.22 \\
FREEDOM & 2.57 & 2.13 & 5.65 & 3.34 & 6.29 & 4.15 \\
MMSSL & 6.31 & 3.77 & 14.67 & 5.34 & 17.04 & 5.81 \\
LGMRec & 4.19 & 2.41 & 8.38 & 3.67 & 9.72 & 4.81 \\
DiffMM & 9.45 & 4.23 & 18.61 & 5.99 & 23.85 & 6.54 \\
\midrule
\textbf{HPMRec} & 5.86&  1.97&  15.80&  3.69&  13.06& 4.51\\
\bottomrule
\end{tabular}
\end{table}

\subsection{Hyper-parameter Analysis (RQ5)}
\label{Sec:hyperparameter}
To evaluate the hyper-parameter sensitivity of HPMRec, we conduct comprehensive experiments on four datasets under varying hyper-parameters settings: \textbf{Algebra Component Number $2^{n+1}$}, \textbf{GCN Layer Number $L$}, \textbf{Regularization Balancing Hyper-parameter $\lambda$}, and \textbf{Self-supervised Learning Balancing Hyper-parameter $\lambda_s$}. The best result of each line is marked in Figure~\ref{fig:algebra_n}-\ref{fig:lambda_s}. According to these results, we have the following observations:

\subsubsection{Performance Comparison $w.r.t$ $2^{n+1}$}
\begin{figure}[h]
    \centering
    \includegraphics[width=1\linewidth]{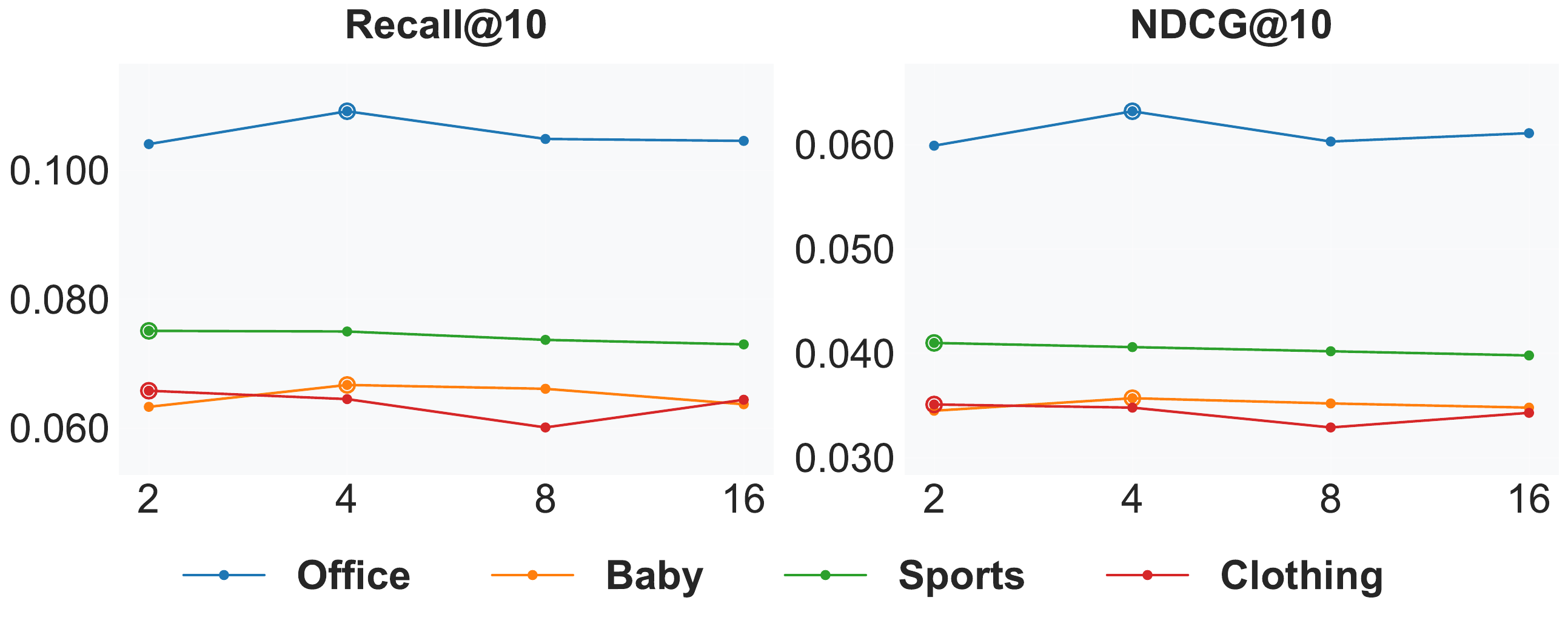}
    \caption{Effect of Algebra Component Number $2^{n+1}$.}
    \label{fig:algebra_n}
\end{figure}

We analyze how different the component number $2^{n+1}$ influences the performance of the HPMRec. According to the result in Figure~\ref{fig:algebra_n}, we found that when component number $2^{n+1}$ equal to 4 ($n=1$), HPMRec achieves optimal performance in terms of Recall@10 and NDCG@10 across Office and Baby datasets, and it achieves optimal performance in terms of Recall@10 and NDCG@10 when component number $2^{n+1}$ equal to 2 ($n=0$) across Sports and Clothing datasets. When the component number is larger than 4 ($n>1$), the performance does not improve, but rather has a negative influence. We attribute this situation to: four components are sufficient for the multimodal information encoder, a larger component number means higher diversity, which might introduce noise, resulting in suboptimal performance.

\subsubsection{Performance Comparison $w.r.t$ $L$}
\begin{figure}[h]
    \centering
    \includegraphics[width=1\linewidth]{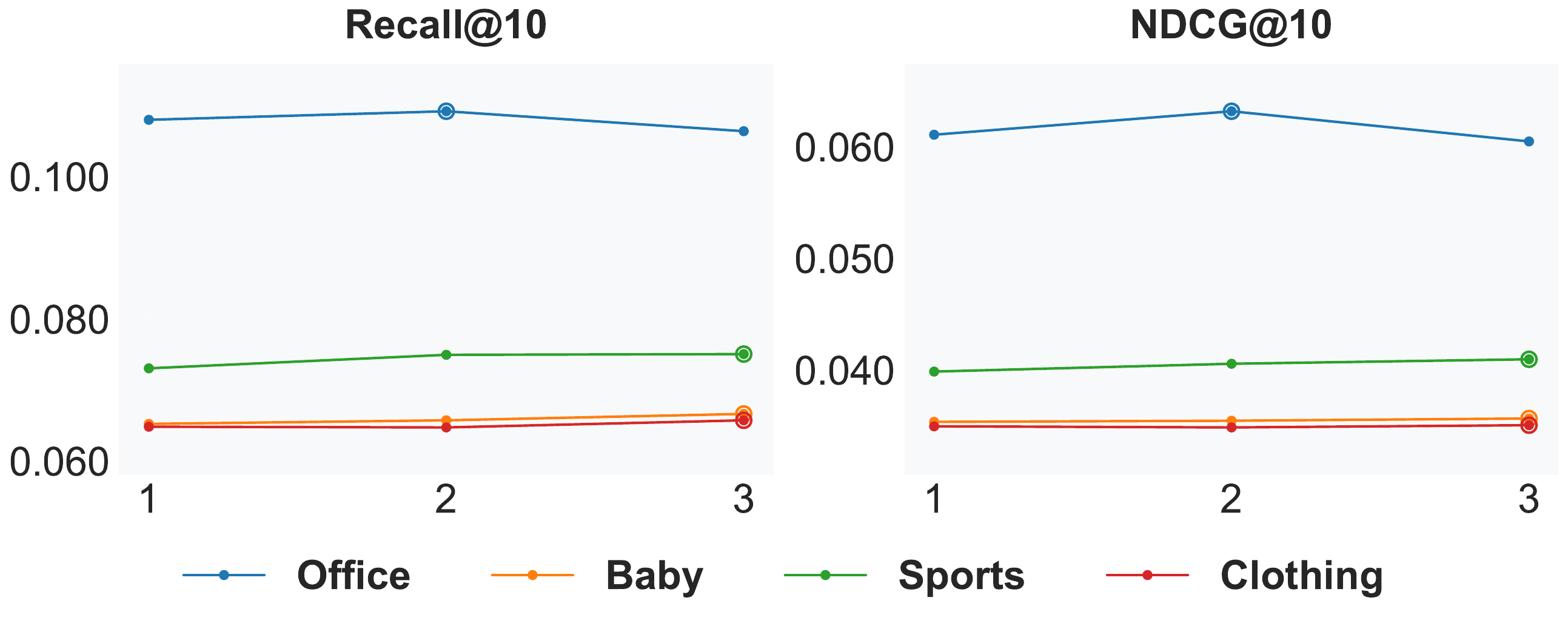}
    \caption{Effect of GCN Layer Number $L$.}
    \label{fig:layer_L}
\end{figure}

As results show in Figure~\ref{fig:layer_L}, we observe that the optimal value of layer number $L$ is different across datasets: In terms of Recall@10 and NDCG@10, the framework achieves the best performance at 3 on the Baby, Sports, and Clothing datasets, and 2 on the Office dataset. Compared to other datasets, the Office dataset has a lower sparsity of user-item interaction, which means shallower GCNs are enough to extract the latent relationship, and the shallower message passing can avoid noise amplification.

\subsubsection{Performance Comparison $w.r.t$ $\lambda$ and $\lambda_s$}
\begin{figure}[h]
    \centering
    \includegraphics[width=1\linewidth]{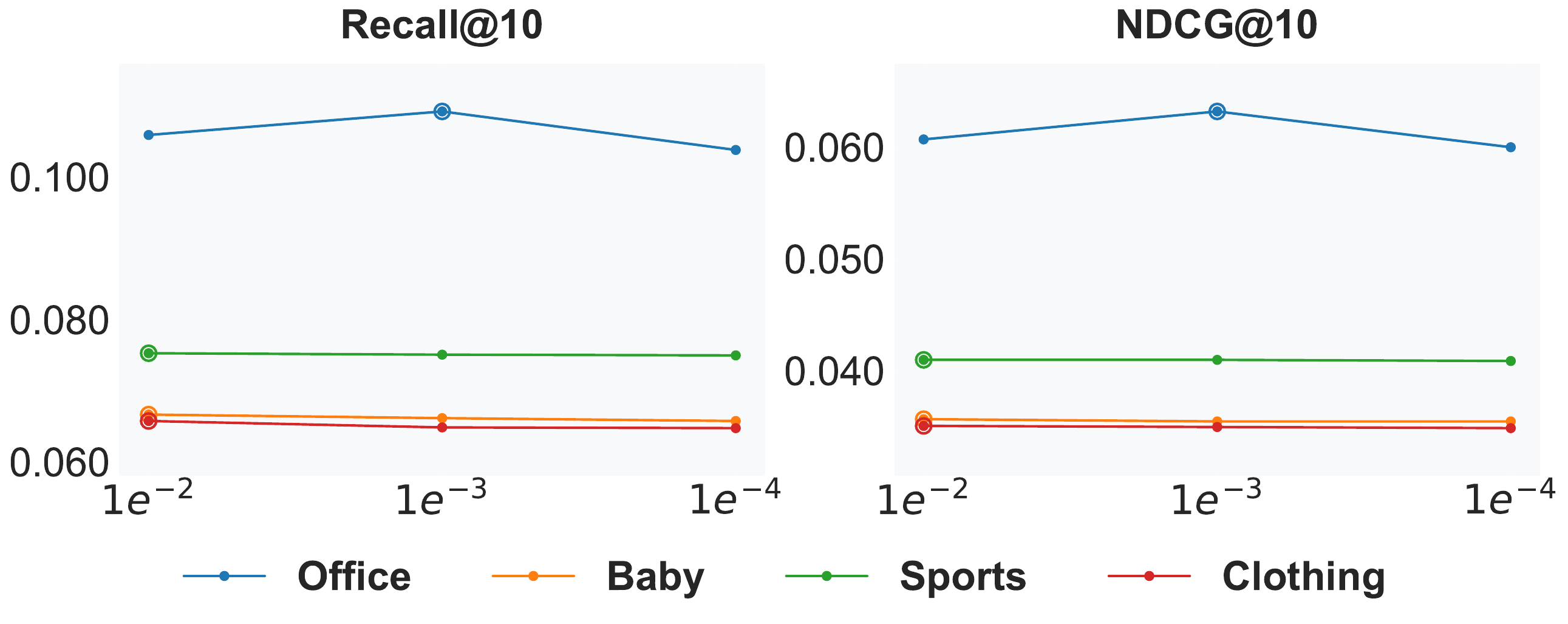}
    \caption{Effect of Balancing Hyper-parameter $\lambda$.}
    \label{fig:lambda}
\end{figure}

\begin{figure}[h]
    \centering
    \includegraphics[width=1\linewidth]{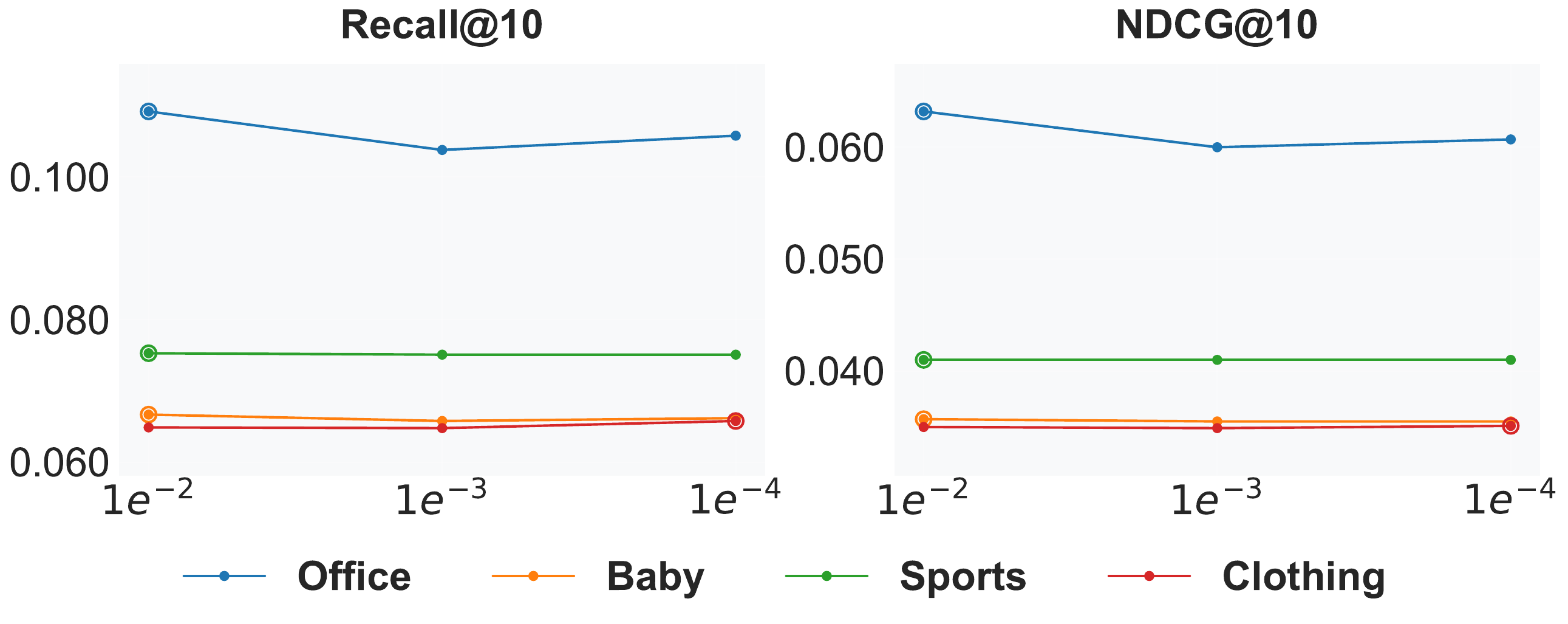}
    \caption{Effect of Balancing Hyper-parameter $\lambda_s$.}
    \label{fig:lambda_s}
\end{figure}

We analyze the effect of the regularization balancing hyper-parameter $\lambda$ (shown in Figure~\ref{fig:lambda}). In terms of Recall@10 and NDCG@10, HPMRec achieves the best performance at $1e^{-2}$ on Baby, Sports, and Clothing datasets, and for Office, $1e^{-3}$ is best. As for the results of self-supervised learning regularizer $\lambda_s$ shown in Figure~\ref{fig:lambda_s}, we find the same optimal setting ($1e^{-2}$) for all other three datasets except for the Clothing dataset. For the Clothing dataset, $1e^{-4}$ is best.


In summary, being flexible in choosing the hyper-parameter settings will allow us to adapt our model to multiple datasets. Although the optimal setting of these hyper-parameters varies, the performance differences are minimal, demonstrating the robustness and stability of HPMRec on different datasets.

\section{Discussion}
Based on the model implementation description in Section~\ref{Sec:methodology} and the results analysis of comprehensive ablation experiments in Section~\ref{Sec:ablation study}, we have the following discussion and interpretation of the effectiveness of each module of HPMRec and our design principles.

\subsection{Model Joint Optimization}
As shown in Section~\ref{Sec:methodology}, our HPMRec adopts multiple modules to jointly optimize. The result analysis in Section~\ref{Sec:ablation study} shows that each components have a positive effect on HPMRec. We will further discuss the crucial synergy and mutual constraints between modules that influence the model optimization.
The prompt-aware compensation mechanism keeps the core modality-specific feature. Meanwhile, real-imag discrepancy expansion task enhances the ability of user/item representation to mine more modality-specific features, which can enhance the diversity of representation. These two modules' mutual constraints ensure that the representation will not lose the core modality-specific features in the pursuit of diversity, and deviate from the reasonable semantic space.
However, the high diversity of each modality's representation will increase the gap between different modalities. To align different modalities and benefit the modality fusion, we design the cross-modality alignment task and MI enhancement fusion strategy. These two modules' synergy ensures that the gap between modalities does not affect modality fusion.
In general, thanks to the prompt-aware compensation mechanism, the representation of each modality retains the core modality-specific features while mining more modality-specific features under the optimization of the real-imag discrepancy expansion task. With the joint optimization of all modules, HPMRec achieves state-of-the-art performance.

\subsection{Maximize the Ability of Prompt}
\label{Sec:maximize}
Through the analysis of variant \textbf{HPMRec $w$ Explicit} in Section~\ref{Sec:ablation study}, we found that the learnable prompt can perform better in implicit guidance than explicit guidance. We attribute this phenomenon to the unsuitable optimization task and insufficient utilization of the powerful dynamic optimization capabilities of the learnable prompt. In our framework HPMRec, the main recommendation task and the self-supervised learning tasks implicitly optimize the prompt to facilitate the learning of core modality-specific features while avoiding the introduction of modality differences, and ensure a sufficiently flexible feature space (solution space)~\cite{fang2023universal} to enhance user/item representations. Therefore, when there are no suitable explicit optimization task, utilizing implicit optimization tasks to ensure a sufficiently solution space of the learnable prompt can maximize its ability. 

\section{Conclusion}
In this paper, we propose HPMRec, a hypercomplex, prompt-aware multimodal recommendation framework that enriches feature diversity and bridges semantic gaps across modalities. Specifically, HPMRec encodes each modality into a multi-component hypercomplex embedding, leveraging the multi-component representation ability of hypercomplex algebra to capture diverse modality-specific features. Secondly, HPMRec leverages the hypercomplex multiplication as naturally nonlinear fusion between modality pairs, thereby exploring more latent cross-modality features. Moreover, to mitigate component misalignment and keep core modality-specific features, we introduce a prompt-aware compensation mechanism that dynamically compensates each component, and this module also mitigates the over-smoothing problem. Finally, we design self-supervised learning tasks to further assist modality fusion and enhance the diversity of modality features. Extensive evaluations on four public datasets demonstrate that HPMRec outperforms state-of-the-art baselines in recommendation performance.

\section*{Acknowledgment}
This work was supported by the National Natural Science Foundation of China (Grant No: 62072390, and 92270123), and the Research Grants Council, Hong Kong SAR, China (Grant No: 15203120, 15226221, 15209922, and 15210023).

\section*{GenAI Usage Disclosure}
No GenAI tools were used in any stage of the research, nor in the writing.



\balance

\end{document}